# Preparation of facilities for fundamental research with ultracold neutrons at PNPI

A.P. Serebrov[1*], V.A. Mityuklyaev[1], A.A. Zakharov[1], A.N. Erykalov[1], M.S. Onegin[1], A.K. Fomin[1], V.A. Ilatovskiy[1], S.P. Orlov[1], K.A. Konoplev[1], A.G. Krivshitch[1], V.M. Samsonov[1], V.F. Ezhov[1], V.V. Fedorov[1], K.O. Keshyshev[2], S.T. Boldarev[2], V.I. Marchenko[2]

[1] *Petersburg Nuclear Physics Institute, RAS, 188300, Gatchina, Leningrad District, Russia*

[2] *P.L. Kapitza Institute for Physical Problems, ul. Kosygina, 2, Moscow 119334, Russia*

[*] *Corresponding author:*
A.P. Serebrov
Petersburg Nuclear Physics Institute
Gatchina, Leningrad district
188300 Russia
Telephone: +7 81371 46001
Fax: +7 81371 30072
E-mail: serebrov@pnpi.spb.ru

**Abstract**

The WWR-M reactor of PNPI offers a unique opportunity to prepare a source for ultracold neutrons (UCN) in an environment of high neutron flux (about $3 \cdot 10^{12}$ n/cm$^2$/s) at still acceptable radiation heat release (about $4 \cdot 10^{-3}$ W/g). It can be realized within the reactor thermal column situated close to the reactor core. With its large diameter of 1 m, this channel allows to install a 15 cm thick bismuth shielding, a graphite premoderator (300 dm$^3$ at 20 K), and a superfluid helium converter (35 dm$^3$). At a temperature of 1.2 K it is possible to remove the heat release power of about 20 W. Using the $4\pi$ flux of cold neutrons within the reactor column can bring more than a factor 100 of cold neutron flux incident on the superfluid helium with respect to the present cold neutron beam conditions at the ILL reactor. The storage lifetime for UCN in superfluid He at 1.2 K is about 30 s, which is sufficient when feeding experiments requiring a similar filling time. The calculated density of UCN with energy between 50 neV and 250 neV in an experimental volume of 40 liters is about $10^4$ n/cm$^3$. Technical solutions for realization of the project are discussed.

## 1. Introduction, evolution of UCN sources and prospects of its future development

Ultracold neutrons (UCN) are successfully used in fundamental research: for the search of the neutron electric dipole moment, for neutron lifetime measurements, for measurement of neutron decay asymmetries and other studies. The accuracy one may reach in these experiments is severely limited by counting statistics. Therefore there is a strong activity in the development of more intense UCN sources (Fig. 1).

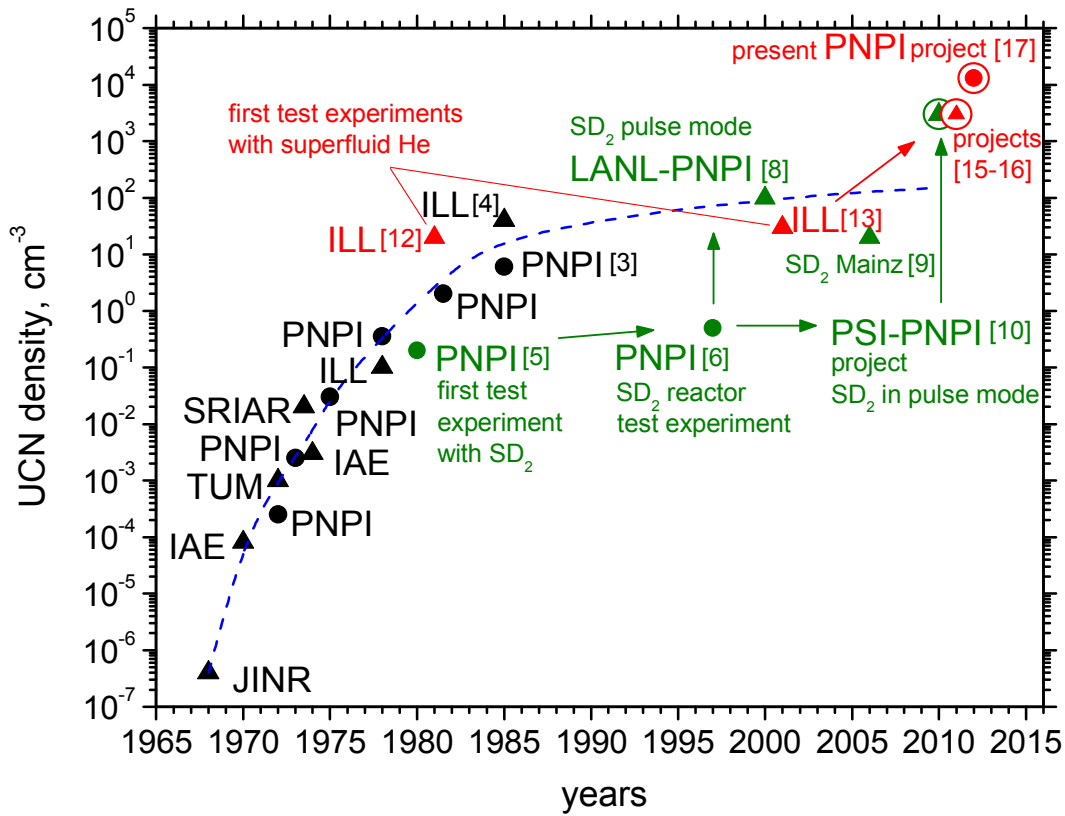

Fig. 1. Development of UCN sources and future prospects. (More detailed information can be found in references [3-6,8-10,12-13,15-17].

The first activity for UCN production was started in Dubna [1] and in Munich [2]. Then UCN density has been increased by eight orders of magnitude due to using of more powerful reactor and cold neutron sources in Gatchina (PNPI) [3] and in Grenoble (ILL) [4]. These sources were placed in extremely high neutron fluxes and used liquid hydrogen (PNPI) or liquid deuterium (ILL). In 90s this line of UCN sources development came to saturation. Development of alternative branches of UCN sources is connected with using of superfluid He 4 at the temperature about 1 K and solid deuterium at the temperature 4 K. The both possible branches of future UCN sources development seem the promising ones.

The first test experiment with solid deuterium has been carried out in Gatchina (PNPI) in 1980 [5]. More detailed studies of solid deuterium UCN source were realized in Gatchina at WWR-M reactor in 1994 [6]. The solid deuterium UCN source with volume 6 liters was installed in thermal column of reactor. It was demonstrated that solid deuterium gives the additional gain factor about 10 times with respect to liquid deuterium. Unfortunately solid deuterium can be used at lower heat release density or in pulse mode when the average heat release is low enough. The pulse mode can be realized at neutron spallation sources therefore the project of so named UCN factory was proposed in [7]. The 1st UCN workshop devoted to this project was organized in Russia (Pushkin) in 1998. The first and simplest realization of this idea was done in LANL [8] and then in Mainz [9]. More detailed project of UCN factory was developed by PNPI and PSI [10] which is now under construction in PSI.

Another branch of UCN sources is connected with using of superfluid He 4 proposed in [11]. The first test experiment with superfluid He was realized in ILL [12] in 1980 and then it was repeated in 2002 [13]. The similar test experiment with superfluid He was done in Japan in 1992 [14]. Now there are a few projects based on the using of superfluid He with UCN accumulation in converter (ILL) [15], or in the regime of steady state (Japan) [16], (Gatchina) [17].

## 2. Scheme of UCN source in thermal column of WWR-M reactor

In this article we discuss the project of a UCN source at the WWR-M reactor of PNPI [14], for which superfluid helium shall be used as a converter for UCN. The thermal column with diameter 1 m, situated close to the reactor core (see Fig. 2), offers a unique opportunity to prepare a source for ultracold neutrons (UCN) in an environment of high neutron flux (about $3 \cdot 10^{12}$ n/cm$^2$/s) at still acceptable radiation heat release (about $4 \cdot 10^{-3}$ W/g). In order to reduce the heat release from γ-rays from the reactor core a bismuth shielding will be installed. The external diameter of this cylindrical shielding will be 990 mm with a wall thickness of 95 mm. The closing of the cylinder in direction to the reactor core will have a thickness of 150 mm. The bismuth is enclosed in a water-cooled aluminium shell (5 mm thick). Inside the bismuth shielding a graphite moderator with thickness 150 mm will be installed. It will be cooled down to 20 K by means of a helium refrigerator. The graphite is also enclosed in an aluminium shell. This construction is placed in a vacuum jacket. Inside the graphite moderator a cylindrical vessel with superfluid helium at a temperature of 1.2 K is placed. Its diameter is 300 mm, and the length is 500 mm. The thickness of the aluminium walls is 2 mm. The internal surface of Al walls is coated by $(3-5) \cdot 10^3$ Å of $^{58}$NiMo alloy with critical velocity 7.8 m/s. UCN can be extracted from the source by means of a UCN guide coated by $^{58}$NiMo, too. UCN guide and source volume will be separated by a 100 µm thick Al membrane with support grid. The scheme in Fig. 2 is shown in scale. The thickness of Bi shielding and graphite premoderator has been chosen to obtain the maximal possible neutron flux with wave length 9 Å, at the condition that the heat load on the superfluid source will not exceed 20 W.

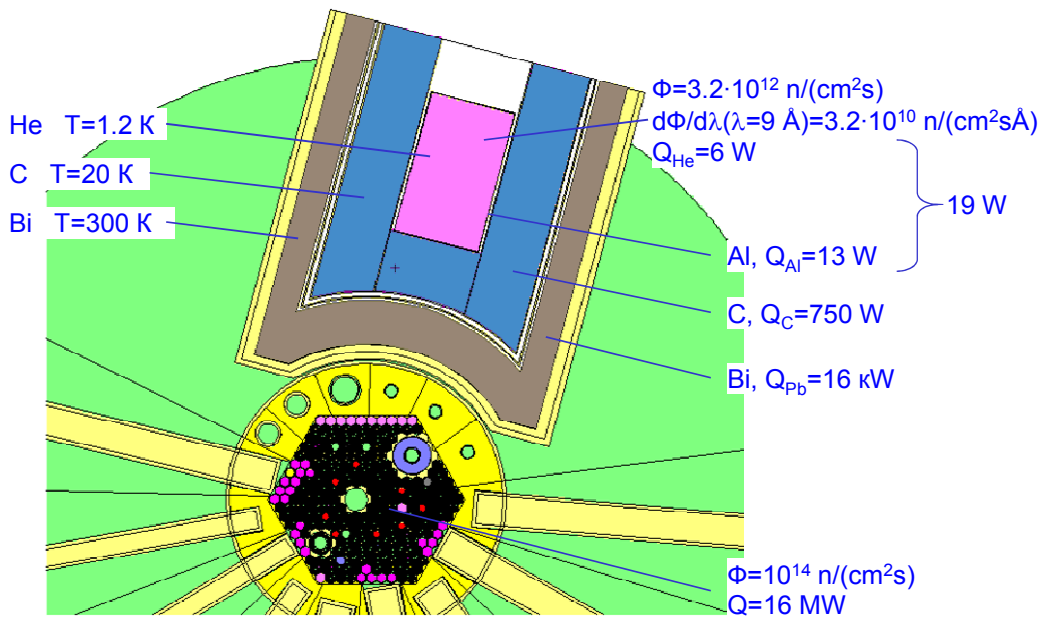

Fig. 2. Scheme of implementation of the planned UCN source into the thermal column of the WWR-M reactor (see text for description of the components).

Calculations of neutron fluxes and the heat release were done using the MCNP code. The total heat releases at 16 MW reactor power are: 16 kW in the bismuth shielding, 750 W in the graphite moderator, 13 W in the aluminium shell of the helium source, and 6 W in the superfluid helium. Hence the total heat load at temperature level 1.2 K is 19 W.

## 3. Premoderator

The choice of graphite as a cold moderator is a compromise between simplicity and the goal to obtain the maximum possible flux of cold neutrons with wave length of 9 Å. These neutrons can be converted into UCN by means of a one-phonon process. UCN production due to multi-phonon processes can be similarly strong but it depends on the neutron spectrum [18]. Fig. 3 shows the results of calculations of the cold neutron spectrum for the following materials: liquid deuterium, graphite, frozen heavy water and beryllium. For all cases the temperature of moderator was chosen to be 20 K. From the practical point of view the graphite moderator is the most simple. However, it is inferior to liquid deuterium in the production of 9 Å neutrons by a factor of two.

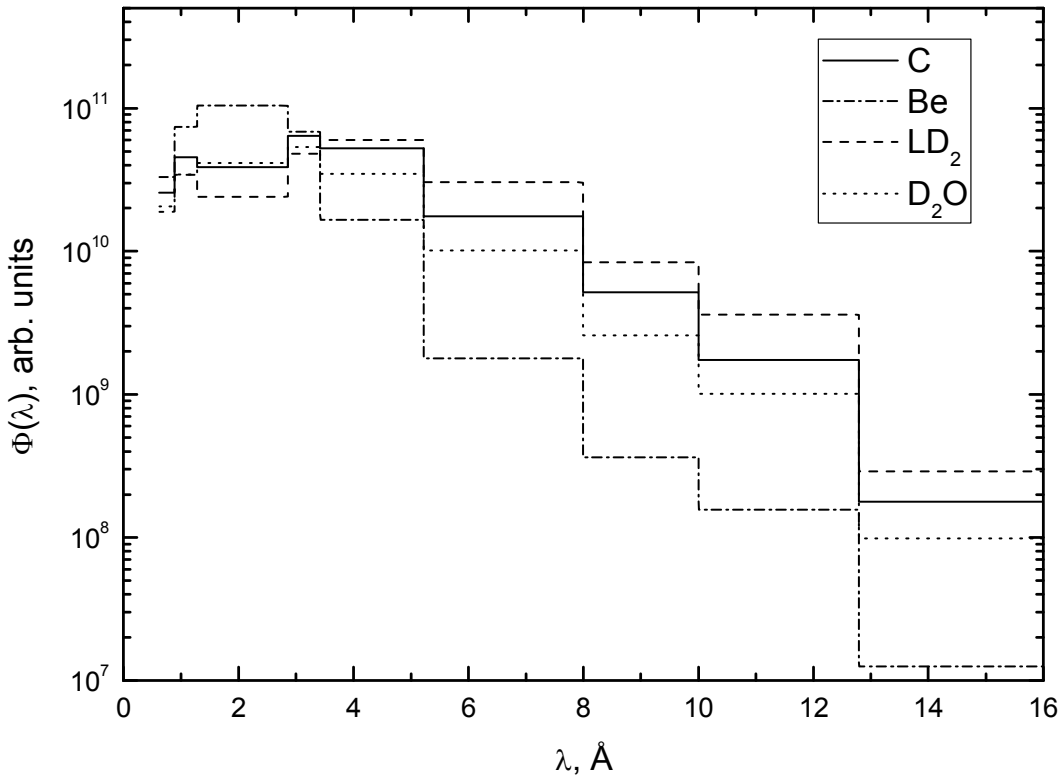

Fig. 3. Neutron spectrum incident on the helium converter, for different premoderators.

## 4. Cryogenic scheme of source

Now we would like to discuss the practical realization of the proposed scheme (Fig. 4). The method of the heat removal is based on the huge thermal conductivity of superfluid helium. The superfluid component is moving to the source of heat and the normal component is moving to the heat exchanger, so it is not necessary to specially arrange the circulation of liquid helium. Superfluid helium around the UCN guide is used as a guide of heat release. Its cross section is 630 cm$^2$, and the length is about 3 m. The difference of the temperature along superfluid helium will be $1.2 \cdot 10^{-2}$ K at a heat load of 20 W. Taking into account a jump of temperature on the surface of the heat exchanger we estimate that the temperature difference between the helium in the source and in the pumping bath will be $2 \cdot 10^{-2}$ K. The temperature in the pumping bath is kept at 1.2 K, corresponding to a pressure of 0.6 mbar. A heat load of 20 W corresponds to a consumption of liquid helium of about 38 l/h. For the pumping of gaseous helium two vacuum pumps with a pumping rate of 8 m$^3$/s will be required. It corresponds to pumping 7.5 l/s of helium gas at atmospheric pressure. These estimations show that the proposed project can be realized.

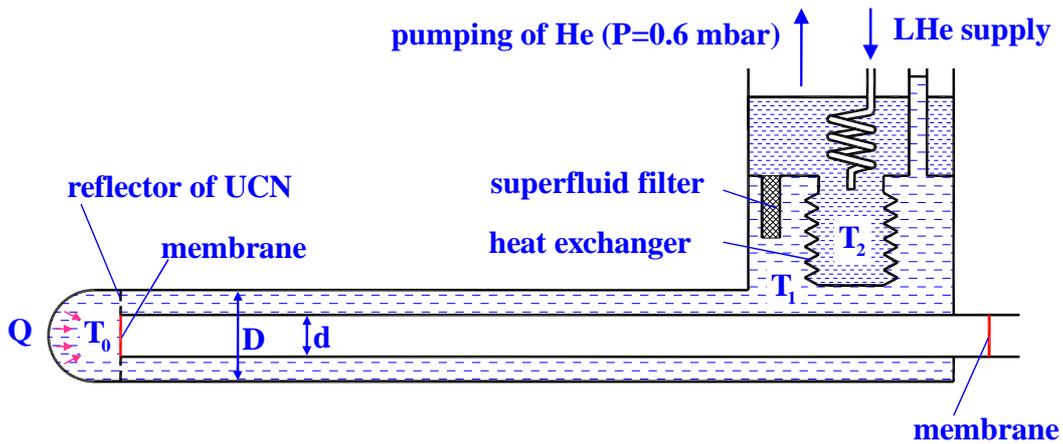

Fig. 4. Cryogenic scheme of the UCN source.

## 5. Estimation of UCN density

Next we estimate the possible UCN density in superfluid helium. MCNP calculations gave the following results for full reactor power of 16 MW. The neutron flux incident on the superfluid helium is $3.2 \cdot 10^{12}$ n/cm$^2$/s, and the differential neutron flux around 9 Å is $3.2 \cdot 10^{10}$ n/cm$^2$/s/Å. The rate $R$ of UCN production due to the one-phonon process can be calculated by the theoretical formula [13] as $R = 4.55 \cdot 10^{-8}$ d$\Phi$/d$\lambda$ (9 Å) cm$^{-3}$s$^{-1}$. Taking into account the multi-phonon processes and neutron spectrum in the source [18] we expect that rate of the UCN production will be $2.9 \cdot 10^3$ n/cm$^3$/s. The total number of UCN produced in the source per second is about $1.0 \cdot 10^8$ n/s. The UCN lifetime in superfluid helium at a temperature of 1.2 K is about 30 s, but after taking into account the UCN losses on the trap walls, it is about 20 s. This estimation was done using a rather pessimistic assumption of $10^{-3}$ losses per collision. Thus the UCN density in a closed trap with superfluid helium could reach $5.8 \cdot 10^4$ n/cm$^3$. However, we are interested in the UCN density in an experimental trap placed in the experimental hall. For this purpose Monte-Carlo calculations have been done for an experimental scheme which includes a trap with superfluid helium with a volume of 35 liters, a UCN guide with diameter 140 mm and length 3 m, and an experimental trap with a volume of 35 liters. In addition, a scheme with an experimental trap of 350 liters has been calculated, too. The volume with superfluid helium is separated from the UCN guide by an aluminium membrane with thickness 100 µm, on a supporting grid. The vacuum of the guide for cold neutrons is also isolated from the warm UCN trap by means of a similar membrane to avoid freezing of residual gas on the cold guide and the source window. The calculations take into account the losses in the walls of the UCN guide and the experimental traps, which were chosen as $3 \cdot 10^{-4}$ per collision. A mirror reflectivity of 99.3% for the guides was used, which was obtained from measurements of the transmission factor for replica guides [19]. The critical velocity of the helium trap and the UCN guide was 7.8 m/s (Ni$^{58}$Mo), but for the experimental trap it was chosen to be 6.8 m/s (Be, BeO). Losses in the aluminium membranes were taken as defined by the capture cross section. The coefficient of UCN reflection from the aluminium membranes was calculated taking into account the critical velocity 3.2 m/s and albedo reflection due to the inhomogeneity of the density inside the foils [20]. It is well-known that reflection plays an important role and can strongly reduce the transmission coefficient. The calculations show that the UCN density in the 35 l experimental trap is by 4.5 times less than the hypothetic UCN density in the closed helium trap, i.e. $\rho_{trap\ 35\ l}$ =$1.3 \cdot 10^4$ n/cm$^3$. The time of filling of the trap in this scheme is 22 s. For the experimental trap with a volume of 350 l the UCN density will be $\rho_{trap\ 350\ l}$ =$7.7 \cdot 10^3$ n/cm$^3$. The filling time constant of the big trap is 55 s. In principle this UCN density can be increased by a factor of 3 by means of changing the configuration of the reactor active core but the heat load will be increased correspondingly. Certainly obtaining a UCN density in the order $10^4$ n/cm$^3$ is very important for fundamental physics experiments with UCN.

## 6. Plan of realization and conclusion

Presently the first steps for the realization of this project have been done. A 3 kW refrigerator with a temperature of 20 K for the cooling of the graphite moderator was put in operation in cryogenic laboratory. In 2009 we have to put in operation helium liquefier which has already arrived in PNPI. The time table of the project strongly depends on the program of the thermal column modernization that will require 3-5 years. Thus an optimistic time estimate of source launching is 2011-2012. This neutron source will produce cold and very cold neutrons as well. The general scheme of the neutron guide halls is presented in Fig. 5. The realization of this project will create a UCN source with record UCN density for fundamental studies. In particular the accuracy of the neutron EDM measurement can be increased by more than an order of magnitude. Besides, cold and very cold neutrons can be used for studies of nanostructures and other research in the field of condensed matter.

This work has been carried out with the support of the RFBR grant 07-02-00859.

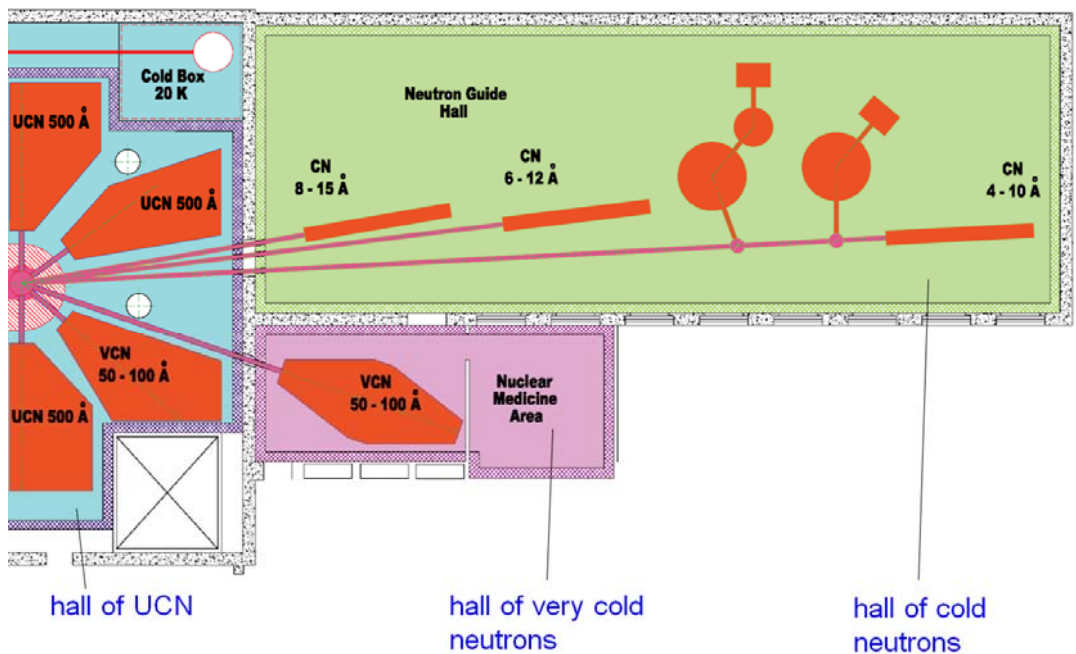

Fig. 5. Neutron guide halls and beam positions around the new source of cold and ultracold neutrons at the WWR-M reactor.